\documentclass{bioinfo}
\copyrightyear{2014}
\pubyear{2014}

\newcommand{\kdet}{{\sc kdetrees}}
\newcommand{\pmc}{{\sc Phylo-MCOA}}

\usepackage[ruled]{algorithm2e}
\SetAlFnt{\footnotesize}
\SetAlCapNameFnt{\footnotesize\relax}
\SetAlCapFnt{\footnotesize\relax}

\usepackage{xr}
\begin{document}
\firstpage{1}

\title[kdetrees: Estimating Phylogenetic Tree Distributions]{kdetrees:
  Nonparametric Estimation of Phylogenetic Tree Distributions}

\author[Weyenberg \textit{et~al}]{%
Grady Weyenberg\,$^{1}$, 
Peter M. Huggins\,$^{2}$,
Christopher L. Schardl\,$^{3}$,
Daniel K. Howe\,$^{4}$,
and Ruriko Yoshida\,$^{1,}$\footnote{to whom correspondence should be addressed}}
\address{
    $^1$Department of Statistics, University of Kentucky, Lexington, KY, USA;\\
    $^2$Robotics Institute, Carnegie Mellon University, Pittsburgh, PA, USA;\\
    $^3$Plant Pathology Department, University of Kentucky,  Lexington, KY, USA;\\
    $^4$Department of Veterinary Science, University of Kentucky, Lexington, KY, USA.}

\history{Received on XXXXX; revised on XXXXX; accepted on XXXXX}

\editor{Associate Editor: XXXXXXX}

\maketitle

\begin{abstract}

\section{Motivation:} 
While the majority of gene histories found in a clade of organisms are
expected to be generated by a common process (e.g. the coalescent
process), it is well-known that numerous other coexisting processes
(e.g. horizontal gene transfers, gene duplication and subsequent
neofunctionalization) will cause some genes to exhibit a history quite
distinct from those of the majority of genes. Such ``outlying'' gene
trees are considered to be biologically interesting and identifying
these genes has become an important problem in phylogenetics.

\section{Results:}
We propose and implement \kdet{}, a nonparametric method of estimating
distributions of phylogenetic trees, with the goal of identifying
trees which are significantly different from the rest of the trees in
the sample.  Our method compares favorably with a similar
recently-published method, featuring an improvement of one polynomial
order of computational complexity (to quadratic in the number of trees
analyzed), with simulation studies suggesting only a small penalty to
classification accuracy.  Application of \kdet{} to a set of
Apicomplexa genes identified several unreliable sequence alignments
which had escaped previous detection, as well as a gene independently
reported as a possible case of horizontal gene transfer.  We also
analyze a set of {\it Epichlo\"e} genes, fungi symbiotic with grasses,
successfully identifying a contrived instance of paralogy.

\section{Availability:}
Our method for estimating tree distributions and identifying outlying
trees is implemented as the {\tt R} package \kdet{}, and is available
for download from CRAN.

\section{Contact:}
\href{ruriko.yoshida@uky.edu}{ruriko.yoshida@uky.edu} 
\end{abstract}

\section{Introduction}
A central problem in systematic biology is the reconstruction of the
evolutionary history of populations and species from numerous gene
trees with varying levels of discordance
\citep{Brito2009,Edwards2009}. Although there is a well-established
understanding that discordant phylogenetic relationships will exist
among independent gene trees drawn from a common species tree
\citep{Pamilo1988,Takahata1989,Maddison1997}, phylogenetic studies
have only recently begun to shift away from single-gene and
concatenated-gene estimates of phylogeny in favor of multi-locus
methods \citep{Carling2008}. These newer approaches focus on the role
of genetic drift in producing patterns of incomplete lineage sorting
and gene tree/species tree discordance, largely using coalescent
theory \citep{Rosenberg2002,Rosenberg2003,Degnan2005}. These
theoretical developments have been used to reconstruct species trees
from samples of estimated gene trees
\citep{Maddison2006,Carstens2007,Edwards2007,Mossel2010,
  RoyChoudhury2008}.

Detecting concordance among gene trees is also a topic of interest.
For example, \cite{bucky} developed a Bayesian method to estimate
concordance among gene trees using molecular sequence data from
multiple loci.  The method can produce estimated gene trees as well as
an estimate of the proportion of the genome that {supports} a particular
clade. However, {\em a priori} assumptions must be made about the
degree and structure of concordance present in the gene trees.

Although there is a tremendous amount of ongoing effort to develop
better parametric models for gene tree distributions, the parametric
framework has inherent limitations. While a parametric method
typically makes the most efficient use of a given data set when the
model is specified correctly, they achieve this efficiency by assuming
that the true distribution of gene trees is one of a relatively small
class of distributions. This can lead to erroneous inferences when
{the} true distribution does not resemble any of the
models in the 
proposed class. Given that many questions remain about the proper way
to incorporate a number of important processes into a parametric model 
(e.g. geographic barriers to migration, or a population bottleneck),
the problem of model mis-specification is very real. Nonparametric
methods avoid the majority of these modeling issues, enabling unbiased
estimation for a much larger class of true tree distributions at a
cost of statistical efficiency.   

Numerous processes can reduce the correlation among gene trees.
Negative or balancing selection on a particular locus is expected to
increase the probability that ancestral gene copies are maintained
through speciation events \citep{Takahata}. {Horizontal transfer
introduces divergent gene copies into a different species through
shuffles gene copies among species} via hybridization 
\citep{Maddison1997}. The correlation may also be reduced by naive
sampling of loci for analysis. For example, paralogous gene copies
will result in a gene tree that conflates gene duplication with
speciation. Similarly, sampled sequence data that span one or more
recombination events will yield ``gene trees'' that are hybrids of two
or more genealogical histories \citep{PosadaCrandall2002}. These
non-coalescent processes can strongly influence phylogenetic inference
\citep{PosadaCrandall2002, MartinBurg2002, Edwards2009}. In addition,
\cite{Rivera} showed that an analysis of complete genomes indicated a
massive prokaryotic gene transfer (or transfers) preceeding the
formation of the eukaryotic cell, arguing that there is significant
genomic evidence for more than one distinct class of genes. These
examples suggest that the distribution of eukaryotic gene trees may be
more accurately modeled as a mixture of a number of more fundamental
distributions.

In this paper, we focus on the problem of identifying significant
\emph{discordance} among gene trees, as well as estimating the
distribution of gene trees as a whole. This set of gene trees is
assumed to consist mostly of ``typical'' (or ``non-outlier'') gene
trees, which are assumed to be independently sampled from some
distribution $f$. For example, gene trees have evolved neutrally under
a coalescent process. In addition, there are a smaller number of
``outlier'' gene trees which are sampled from a very different
distribution $f'$. These genes are assumed to arise from less common
evolutionary processes; for example, paralogy, neofunctionalization,
horizontal gene transfer, or periods of rapid molecular evolution. In
addition, more mundane errors---such as incorrect sequencing,
alignment, tree reconstruction, or annotation---can also produce
outlier trees in a data set \citep{Horner}. Our method produces a
{\em nonparametric} estimate of the distribution $f$ and also attempts
to identify potential outlier gene trees which are probably not
generated by $f$. Trees identified as outliers can then be inspected
more closely for biologically interesting properties. In particular,
identifying and removing outliers that violate model assumptions can
improve the accuracy of inferences made from a collection of gene
trees (e.g. \cite{Disotell, MartinBurg2002,
  Edwards2009,PosadaCrandall2002}).  {Note that in this
  paper we use {\em dissimilarity maps}, {\em geodesic distances}, and
  {\em topological dissimilarity maps} between trees for simulations
  and implementation of our software (see Subsection
  \ref{choice}). With these distance measures between trees, we implicitly assume a multispecies
  coalescent model 
 \cite[]{Rosenberg2012}.  Also note that the choice of tree distance measures might change the
detected outlying gene trees.  For example, if the subtree pruning
and regrafting (SPR) distance between trees is used, the detected
outlyng gene trees would be having an excess of recombinations or horizontal
gene transfers.  }

\subsection{Related Work}

The method presented in this paper is not, at its present state of
development, a statistical method for hypothesis testing, but rather
for discovering possible outliers present in a given collection of
orthologous genes. However, there has been significant work devoted to
the development of statistical methods for testing hypotheses of
discordance between the trees in a collection. The reviewed methods in
\cite{Poptsova2009} are the following: (i) likelihood-based tests of
tree topologies, such as
Kishino-Hasegawa test \citep{HK}, Shimodaira-Hasegawa test
\citep{SH1999}, Approximately Unbiased tests \citep{AU}; (ii) tree
distance methods, such as \citet{Robinson1981}  and subtree
pruning and regrafting distances \citep{goloboff2008calculating}; and
(iii) genome spectral approaches, such as bipartition \citep{penny95}
and quartet decomposition analyses \citep{quartet}.

The likelihood-based tests of tree topologies and tree distance
methods are statistical hypothesis tests that detect significant
incongruence between trees, i.e., they are testing the following
hypotheses: \begin{quote}
  $H_0$: Given trees are topologically congruent. \\
  $H_1$: Given trees are topologically incongruent.
\end{quote} 
The distinction between likelihood and distance based methods is in
how they calculate the p-value of these hypotheses. The
likelihood-based tests compare each gene tree with
a species/reference tree using a likelihood value, to see if the
incongruence is ``statistically significant.'' These methods are also
known as partition likelihood support (PLS) \citep{PLS}. Tree
distance methods estimate the p-value of the hypotheses above by
computing a distance between a reference tree and each gene
tree. \cite{Holmes2005} describes a framework for statistical
hypothesis testing on trees based on tree distances using
distributions of phylogenetic trees (e.g. a posterior distribution or
bootstrap resampling). Holmes also presents a statistical method to
compare two sets of bootstrap sampling distributions, using the mean
and variance of each distribution \cite[Section 4.4.1]{Holmes2005}. A
nonparametric method for detecting significant discordance
between two sets of trees via supporting vector machines (SVMs)
was introduced by \cite{svn}. This is a nonparametric method for
statistical testing of the hypotheses:
\begin{equation*} \begin{array}{ll}
    H_0:& \mbox{Two sets of trees are drawn from the same distribution.}\\
    H_1:& \mbox{ Two sets of trees are not drawn from the same
      distribution.} \end{array} \end{equation*}

While likelihood-based tests assume that the species tree is known,
genome spectral approaches do not use such a reference tree. Genome
spectral methods summarize a set of gene trees with phylogenetic
spectra (frequencies), such as splits or quartets. These frequencies
can be used to approximate the distribution of gene trees, instead of
producing a summarizing tree. Outlier trees can be identified by
looking for trees whose highly supported features disagree with
prevalent features in the spectra \citep{scps}.

A non-statistical approach for summarizing collections of gene trees
is presented by \cite{Nye2008}. Treating each gene tree as a leaf
node, a ``meta-tree'' is constructed where nodes correspond to
phylogenetic trees; distances between nodes of the meta-tree
correspond to distances between phylogenetic trees, and internal nodes
correspond to gene trees with various branches collapsed. When using
the Robinson--Fould distance, the nonparametric method proposed in
this paper can be viewed as a numerical summarization of the meta-tree
in \citep{Nye2008}.

Recently, \cite{MCOA} developed a statistical nonparametric method to
detect outlier trees from the set of gene trees. They first convert
gene trees into vectors in a multi-dimensional Euclidean space and
then apply Multiple Co-Inertia Analysis---an extension of Principal
Coordinate Analysis (PCO)---directly to these vectorized gene trees.
Their method, \pmc{}, also detects outlier species, those whose
position varies widely from tree to tree. Included in our results are
simulation studies comparing our nonparametric method with \pmc{}.

\begin{methods}
\section{Methods}
\subsection{Algorithm}
Let ${\mathcal T}_n$ denote the set of all tree topologies (including
multifurcating trees) on $n$ taxa 
(which we call {\em tree space}). We consider trees to be
unrooted, but rooted trees can be treated similarly.  Our main object
of study is a sample, $\{T_i\}_{i=1}^N$, of $N$ trees (gene trees)
mostly drawn from a distribution $f$ on ${\mathcal T}_n$.  If $n$ is
large enough that $|{\mathcal T}_n| \gg N$ then many tree topologies
in the sample may have low empirical frequency. In this case, $f$
cannot be estimated well by assigning $\hat{f}(T)$ to be the empirical
frequency of $T$ in the sample. On the other hand, if $f$ corresponds
to a model such as the coalescent, it is reasonable to expect that
topologies ``close'' to many observed trees will have a higher
likelihood than topologies ``far away'' from the observed trees.

{\em Kernel density estimation} is a nonparametric technique to
estimate a distribution that generated a sample, by leveraging the
fact that points close to sample points tend to have higher likelihood
than distant outlier points (under adequate assumptions on the
distribution, namely, the distribution is square-integrable
\citep{Meloche}). Kernel density estimation can be viewed as a refined
version of histogram-based estimation of a density. 

Given an independent and identically distributed sample of trees $T_1,
\ldots, T_N$, we propose a nonparametric estimator of the distribution
that generated the sample with the form
\[\hat{f}(T) \propto \frac{1}{N} \sum_{i=1}^N k(T, T_i). \]
Here $k$, the kernel function, is a non-negative function defined on
pairs of trees which measures how ``similar'' two trees are. For our
approach, we do not require $k$ to be a kernel in a strict statistical
sense.

In \kdet{} we have implemented a kernel of the form
\[ k(T, T_i) \propto \frac{1}{h_i}\exp \left( { -
    {\left({\frac{d(T,T_i)}{h_i}}\right)}^\delta }\right). \] A
distance function on the space of trees, $d(T,T')$, is used to define
a univariate projection $\mathcal T_n \to \mathbb R_+$ in the natural
way for each fixed $T\in\mathcal T_n$, mapping $T' \mapsto
d(T,T')$. The ``shape'' parameter $\delta > 0$, and the ``bandwidth''
parameters $h_i > 0$ control how tightly each contribution $k(T,T_i)$
will be centered on
$T_i$.  
Allowing the bandwidth to vary with the sample points, $T_i$, is
called an \emph{adaptive bandwidth} method. Alternatively the
bandwidth can be set to a constant value for all $T_i$.

In general, we can remove the symmetry and triangle inequality
requirements for $d$, and it is possible that the sum over tree space,
$\sum_{T\in\mathcal T} k(T, T')$, will vary with $T'$. Ideally, we
would remedy this issue by normalizing $k(\cdot,T')$ so that
$\sum_{T\in\mathcal T} k(T, T') = 1$. (This is the case most analogous
to kernel density estimation.) 
However, for the $d$ implemented by \kdet{}, Monte Carlo estimates of
this sum do not appear to vary significantly across $T'$, and so the
current version of the software assumes that it is
constant. (Additional information about these estimates is presented
in Supplementary Figure \ref{fig:kernalvol}.)


Since the ultimate goal is to detect outlier trees, $T_j$, which are
not actually drawn from the true distribution $f$, we are most
concerned with estimating the density at the observed sample
points. In this context, it makes sense to use a ``leave-one-out''
estimator which excludes the contribution of the point in question
from the tree score,
\[\hat{g}(T_j) = \frac{1}{N-1} \sum_{i \neq j} k(T_j, T_i). \]

Once we have computed the scores, $\{\hat g(T_i)\}$, we classify tree $T_j$
as an outlier if $\hat g(T_j)$ is less than $Q_1 - \kappa \cdot IQR$. 
Where $Q_1$ and $IQR$ are the first quartile and the interquartile
range of the set of tree scores, respectively; and $\kappa$ is a
classification tuning parameter. The choice of $\kappa$ affects the
sensitivity and specificity of the classifier, and is set
to $1.5$ by default { as defined by J. Tukey for finding
  outliers \cite[]{Tukey1977}}, although the user may supply their own value. 
 

\subsubsection{Choice of tree distance:}\label{choice}

In our approach, trees can be incorporated into a statistical
framework by converting them into a numerical vector format based on a
distance matrix or map. These vectorized trees can then be analyzed as
points in a multi-dimensional space where the distance between trees
increases as they become more dissimilar \citep{Hillis:2005xc,
  Semple:2003tp, Graham:2010qa}. 

For the choice of $d$, we propose distances derived from three different
distances on trees: {\em dissimilarity map} $d_{d}$, {\em
  topological dissimilarity map} $d_{t}$, and {\em geodesic
  distance} $d_{geo}$.  The dissimilarity map
distance measure between two trees is the Euclidean distance,
\[d_{d}(T',T) = || v_{d}(T) - v_{d}(T') ||_2, \] where $v_{d}(T)$ is a
vectorization of trees, $\mathcal T_n \to \mathbb R^{n\choose 2}$,
based on an enumeration of the pairwise distances between the tips
\citep{Buneman1971}. The topological dissimilarity map distance
measure between two trees is defined similarly, \[d_{t}(T',T) = ||
v_{t}(T) - v_{t}(T') ||_2, \] but uses a vectorization $v_t(T)$ that
counts the number of edges between the tips \citep{Steel1993}. An
example calculation of both $v_{d}$ and $v_{t}$ is shown in
Supplementary Figure \ref{fig:treevec}.

\cite{billera2001geometry} showed that the space of rooted trees with
a fixed number of taxa is the union of positive cones in ${\mathbb
  R}^{\binom{n}{2}}$. Thus, the space of trees is the set of all
metrics derived from valid trees, and is a subspace of the space of
all distance matrices.  The geodesic distance $d_{g}$ is the shortest
distance between two valid trees when the connecting path is
constrained within this tree space (note that this subspace of valid
trees is not itself Euclidean).  \cite{owen2011fast} developed an
$O(n^4)$ algorithm to compute the geodesic distance $d_{g}(T, T')$
between any two valid trees.


\subsubsection{Missing taxa:}

It is desirable for phylogenetic analyses to be able to deal with
situations with incomplete data. In this case, the most relevant type
of missing data is when some gene trees are missing a tip which is
present in other trees in the data set. Our method is capable of
handling such a situation if the dissimilarity or topological distance
maps are used. In this situation we impute missing tip-to-tip
distances in the tree vectors with the median value found in trees
containing the missing tip. Unfortunately, the geodesic distance
algorithm we employed does not currently allow us to perform such an
imputation, and so \kdet{} cannot handle missing tips if the geodesic
distance map is selected.

If the trees have node labels which correspond to support for the
given split (obtained, for example, by a bootstrap analysis), then the
software can accommodate this information by collapsing nodes with
support less than a given value. This behavior is disabled by default.

\subsubsection{Kernel bandwidth:}

The estimator $\hat{g}$ depends crucially on the choice of the
bandwidth parameter $h$. We employ a nearest-neighbor approach to
estimate an adaptive bandwidth for each sample point. To estimate the
bandwidth for a point $T_j$, we use the distance to the $m$-th closest
sample point. This approach has the effect of causing the kernels to
be concentrated in areas where there is a lot of data, and diffuse in
the tails of the distribution. In the current version of \kdet{} $m$
is defaulted to be 20\% of the sample size, a heuristic value chosen
based on simulation results.

Alternatively, the bandwidth can be set to a constant value for all
$T_i$. In order to do this we must find a way to choose an optimal
value for the bandwidth $h$. We experimented with a constant bandwidth
chosen by estimating the partition function $Z_h = \sum_T
\hat{g}_h(T)$ using a random sample of trees.  However, it seems that
we tend to under-estimate the bandwidth $h$ and the results are not as
robust as in the case of the adaptive bandwidth.

\subsubsection{Tuning parameters:}
The outlier classifier's sensitivity depends on the choice of a
tuning parameter, $\kappa$. The default value, 1.5, is chosen for
historical reasons. In our simulations smaller values of $\kappa$, around 0.75 to
1, often resulted in false positive rates close to 5\%.  Creating plots of the
tree scores may be helpful in choosing an appropriate value for a given
data set.

\subsubsection{Computational complexity:}
The running time of \kdet{} is dominated by the step where
pairwise tree distances are calculated. For $N$ trees, each with $n$
taxa, this step takes $O(n^2N^2)$ operations when using the
dissimilarity or topological distances, or $O(n^4N^2)$ if using the
geodesic distance.

\subsection{Simulations}
We conducted a series of simulations comparing the performance of
\kdet{} and \pmc{}. (Code and documentation for the simulations is
included in a package vignette with \kdet{}.)
The simulated data consisted of coalescent trees generated by the
Python library DendroPy \citep{dendropy}. Six species trees (see
Supplementary Figure \ref{fig:speciestree}) were used to contain
coalescent gene trees. A data set consisted of a small number of
``outlier'' gene trees, together with a larger number of
``non-outlier'' gene trees.
Pseudocode in Algorithm \ref{alg:sim1} summarizes the simulation
processes.

\begin{algorithm}[!tpb]
  \KwIn{Coalescent population parameter. Number of non-outlier trees,
    $g$. Number of random outlier trees, $r$. Set $S$ of species trees. Classification tuning parameter, $\kappa$.} 
  \KwResult{Average number of true and false outlier identifications for each method.}
  \For{each iteration in simulation}{
    Generate the set of non-outlier trees by sampling $g/|S|$
    coalescent gene trees from each $s\in S$\;
    Generate $r$ random outlier gene trees, each within a new random
    species tree\;
    Analyze data with both \kdet{} and \pmc{}\;
    Tally true and false outlier identifications for each method\; }
  \caption{Summary of the simulation comparing \kdet{} and
    \pmc{}. (See Supplementary Figure \ref{fig:speciestree} for a plot
    of the species tree used.)  For the ``single'' simulations, $S$
    contains a single tree (top left of Figure \ref{fig:speciestree}), while for the ``mixed'' simulations it
    contained 5 trees (remainder of Figure \ref{fig:speciestree}). For our simulations, $r=1$ and $g=100$.}
  \label{alg:sim1}
\end{algorithm}


Our first simulation investigated the classification characteristics
of the methods, producing receiver operating characteristic (ROC)
curves comparing \kdet{} and \pmc{}, by varying the classification
tuning parameter of each method. (A ROC curve is a graphical plot of
the fraction of true positive rate vs. the fraction of false positive
rate at various threshold settings \citep{Hastie}.) In this simulation
we set the effective population size of the coalescent process
generating the trees to 2000, a value which produced a moderate amount
of variance in the generated coalescent trees.

A second simulation compares the true positive rates of the methods as
the variance of the coalescent trees increases. (Variance of the
random trees is controlled by the coalescent population parameter.)
This simulation was carried out both with the default classification
tuning values, as well as values chosen based on the ROC simulation
results to limit the false positive rate (FPR) to around 5\%.

A third simulation compared the distribution of outlier tree scores to
the distribution of non-outlier tree scores. 
The simulation process is summarized in the pseudocode in Algorithm \ref{alg:sim2}.

\begin{algorithm}[!tpb]
  \KwIn{Coalescent population parameter. Number of non-outlier trees,
    $g$. Number of outlier trees, $R$.} 
  \KwResult{Estimate of outlier and non-outlier  tree score distributions.} 
  Generate $g$ coalescent trees within one species tree\;
  Use \kdet{} to obtain scores for non-outlier trees\;
  \For{r=1 \KwTo R }{
    Generate a single outlier tree within a new species tree\;
    Append outlier tree to set of non-outlier trees\;
    Obtain and record outlier tree score\;
  }
  Plot kernel density estimates for both score distributions\;
  \caption{Summary of the simulation design for the
    simulation comparing the tree score distributions for outlier
    trees and non-outlier trees. For our simulations both $g$ and $R$
    are set to 500, and the coalescent parameter is 2000.}
  \label{alg:sim2}
\end{algorithm}



\subsection{Biological datasets}
\subsubsection{Apicomplexa:}

The Apicomplexa data set presented by \cite{kissinger} consists of
trees reconstructed from 268 single-copy genes  from the following species:
{\it Babesia bovis} (Bb) \citep{Brayton2007} (GenBank
accession numbers AAXT01000001--AAXT01000013), {\it Cryptosporidium
  parvum} (Cp) \citep{Abrahamsen2004} from CryptoDB.org
\citep{Heiges2006}, {\it Eimeria tenella} (Et) from GeneDB.org
\citep{Hertz-Fowler2004}, {\it Plasmodium falciparum} (Pf)
\citep{Gardner2002} and {\it Plasmodium vivax} (Pv) from PlasmoDB.org
\citep{Bahl2003}, {\it Theileria annulata} (Ta) \citep{Pain2005}
from GeneDB.org \citep{Hertz-Fowler2004}, and {\it Toxoplasma gondii}
(Tg) from Toxo-DB.org \citep{Gajria2008}. A free-living ciliate, {\it
  Tetrahymena thermophila} (Tt) \citep{Eisen2006}, was used as the
outgroup. To this set of sequences, we appended the Set8 gene, which
has been identified by \cite{Kishore} as a probable case of horizontal
gene transfer from a higher eukaryote to an ancestor of the
Apicomplexa.

\subsubsection{Epichlo\"e:}

Another set of biological sequences to use as a test case was
generated from housekeeping genes and a known pair of paralogs in
{\it Epichlo\"e} species and related plant symbionts and parasites in the
fungal family Clavicipitaceae. We previously reported sequencing,
annotation, and the identification of orthologs in genome of {\it Epichlo\"e
amarillans} strain E57, {\it E. brachyelytri} E4804, {\it E. festucae} strains
E2368 and Fl1, {\it E. glyceriae} E277, {\it E. poae} E5819, {\it E. typhina} E8,
{\it Aciculosporium take} MAFF-241224, {\it Claviceps fusiformis} PRL 1980,
{\it C. paspali} RRC-1481, {\it C. purpurea} 20.1, {\it Neotyphodium gansuense} e7080,
and {\it Periglandula ipomoeae} IasaF13 \citep{Schardl2013}. We compiled
the inferred protein sequences for ten housekeeping proteins, namely,
$\gamma$-actin (ActG), DNA lyase (ApnB), a calmodulin-dependent protein
kinase (CpkA), the largest and second largest subunits of RNA
polymerase II (rpbA and rpbB), translation elongation factor 1-$\alpha$
(TefA), $\alpha$-tubulin (paralogs TubB and TubC), and $\beta$-tubulin
(paralogs TubB and TubP). 
{As a possible phylogenetic outlier, we used an alignment of proteins
related to the {\it Emericella nidulans} {\it O}-acetylhomoserine (thiol) lyase
enzyme (CysD). In some but not all of the fungal strains we analyzed,
the CysD homologs were located in the loline alkaloid biosynthesis
gene cluster, and have been designated LolC
\cite[]{Schardl2013}. Analysis by {\tt OrthoMCL} \cite[]{orthomcl} grouped
all of the 
CysD-related proteins as orthologs, whereas further analysis with
{\tt COCO-CL} \cite[]{Jothi} separated LolC from the other
CysD-related sequences as paralogs. }

\end{methods}

\section{Results}
We present the software package \kdet{} for nonparametric
estimation of tree distributions and detection of outlier trees.  The
software takes as input a sample of trees in Newick format, and
estimates for each tree a ``score'' based on a nonparametric estimator
of the tree density. It can then use these scores to identify putative
outlying trees in the sample. The trees scores and summary plots are
produced as output.

The \kdet{} package is written in {\sc R} \citep{r-core}, and depends
on packages {\sc distory} \citep{distory}, {\sc ggplot2}
\citep{ggplot2}, and {\sc ape} \citep{ape}. The software is available
for download from CRAN and is compatible with all systems supported by 
{\sc R}.

\subsection{Simulation Results}
\begin{figure}[!tpb]
\centerline{\includegraphics[width=86mm]{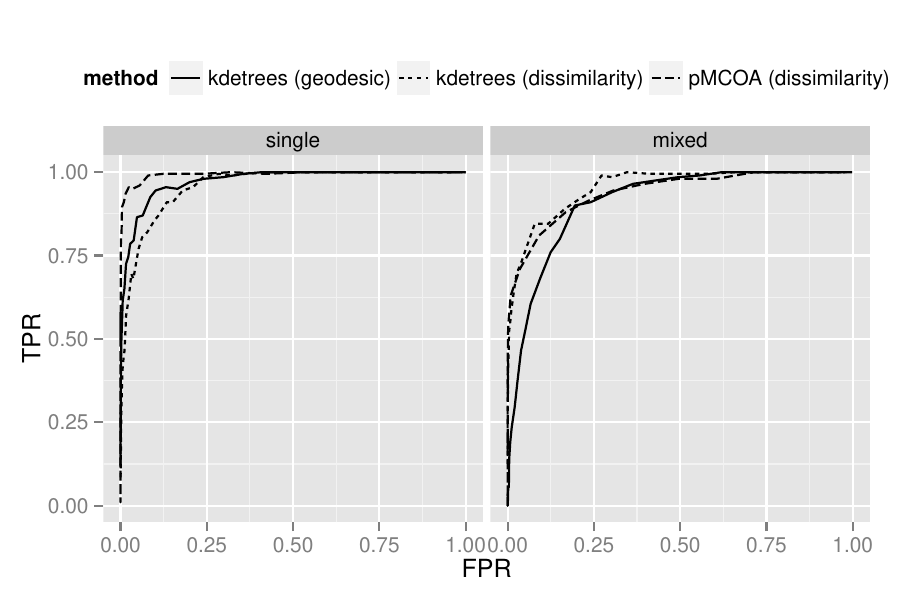}}
  \caption{ROC curves comparing \kdet{} and \pmc{} as
    the classification tuning parameter is varied. (In general higher is better, a very effective classifier will pass close to the upper left corner.) The effective
    population size is 2000 for the coalescent trees.  At left are the
    ``single'' contained coalescent simulations, with the non-outlier
    trees all contained within a single species tree. At right are
    results from a ``mixed'' simulation, with the non-outlier trees
    generated from a mixture of 5 species trees.}
  \label{fig:roc}
\end{figure}

Our first simulation, presented in Figure \ref{fig:roc}, produced ROC
curves comparing the various methods of outlier identification. We
find that the performance of \kdet{} and \pmc{} is similar, with
\pmc{} having a slightly better curve in the single simulations, and
\kdet{} in the mixed scenarios. Interestingly, the geodesic distance
worked better for the ``single'' data than the dissimilarity map,
while the relationship is reversed for the ``mixed'' simulation. These
results were almost completely unaffected by changes in the proportion
of outliers in the sample (proportions between 1 to 10\% were tested).

\begin{figure}[!tpb]
\centerline{\includegraphics[width=86mm]{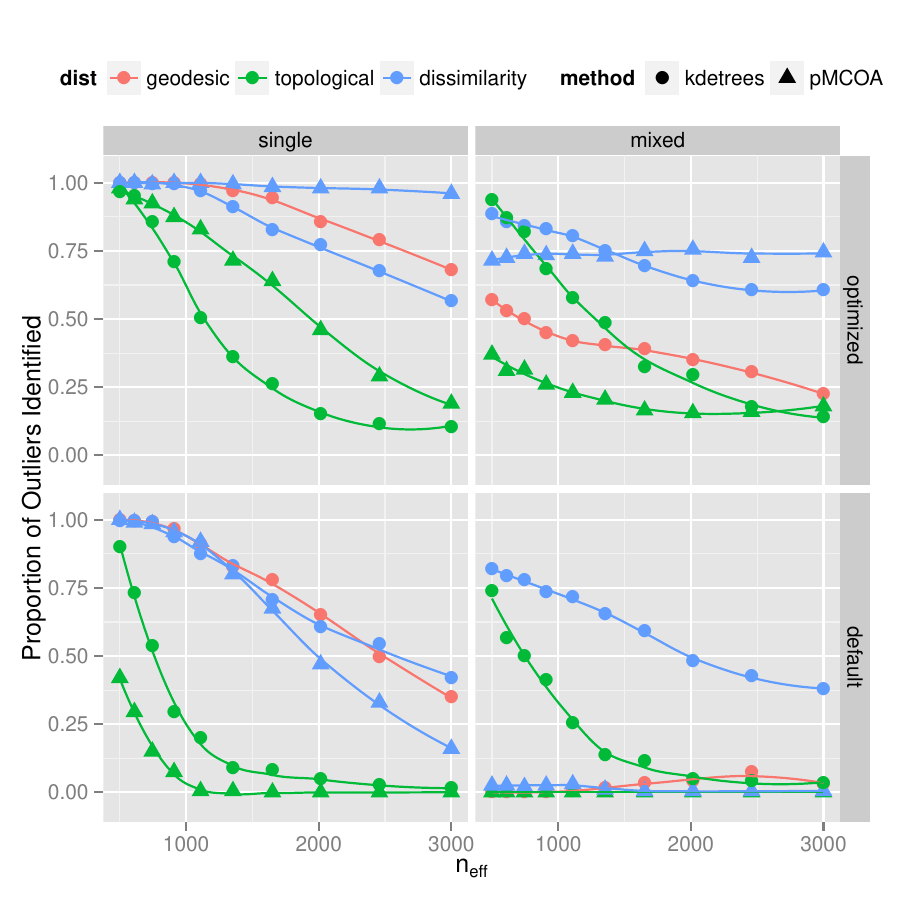}}
\caption{Summary of simulation results comparing performance of
  \kdet{} and \pmc{} for various values of the effective population
  size.  Shown is the proportion of simulated data sets in which the
  methods identified the outlier tree. The top two plots use use
  tuning parameters chosen based on results of the ROC simulation,
  while the bottom plots use default values. For \kdet{} the optimal tuning
  parameter was $\kappa=0.7$, while for \pmc{} it was
  $\kappa=0.25$. The default values are both $\kappa=1.5$.}\label{outliers}
\end{figure}

The variability of the coalescent trees is determined by the effective
population size, the parameter studied in our second simulation. The
proportion of the simulated data sets where each method correctly
identified an added outlier tree is illustrated in Figure
\ref{outliers}. This simulation was run both with default tuning
parameters and ones chosen based on the ROC curve simulation results.
If optimal tuning parameters are selected, \pmc{} can outperform
\kdet{}, however, selecting these correctly can be difficult. 

\begin{figure}[!tpb]
\centerline{\includegraphics[width=86mm]{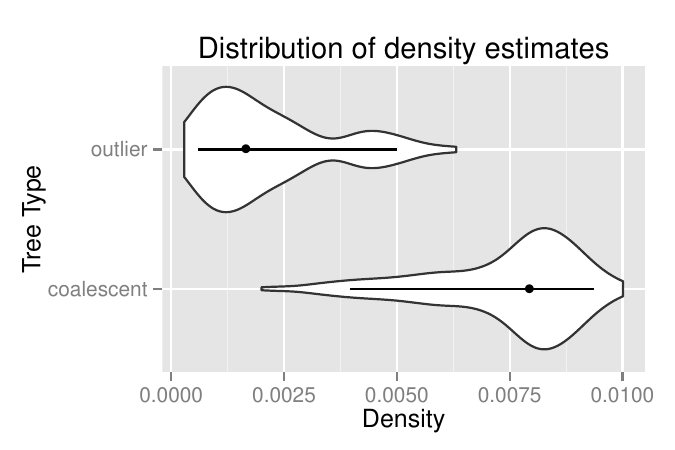}}
\caption{Kernel density estimates of the observed
  distribution of tree scores. The ``coalescent'' scores are for
  contained coalescent trees generated within a fixed species tree
  (bottom). A single random outlier tree is added to this data set and
  its score computed. This process is replicated to generate the
  sample of ``outlier'' tree scores (top). Lines and dots represent
  the 5\%--95\% quantiles and the median, respectively. An effective
  population size of 2000 was used to produce these estimates.}
\label{outliers2}
\end{figure}

We ran a third simulation studying the difference between the score
distributions of outlier trees and non-outlier trees, as the ability of
our method to reliably detect outlying trees depends on a tendency by
outlier trees to produce scores significantly lower than the scores
of non-outlier trees.  The results are presented in
Figure \ref{outliers2}.  We found that while there is some overlap
between the score distributions, the distribution of scores for
outlier trees lies significantly below that of non-outlier
trees. 

\subsection{Biological data results}

\subsubsection{Apicomplexa:}
The list of putative outlier genes identified by \kdet{} in the
Apicomplexa data is presented in Table \ref{names}, with additional
discussion in Supplemental Table \ref{result_summary}. When employing
either the dissimilarity maps or geodesic distance, our method identified
the same set of putative outlier trees. (The first four trees
identified as putative outliers are also plotted in Supplemental
Figure \ref{outTree1}, and the entire set of estimated scores are
summarized in Supplementary Figure \ref{jess}.) These trees all
contain a branch with a length that is far too long in proportion to
the other branches, leading to their identification as
outliers. Closer inspection of these trees suggested that they
correspond to questionable sequence alignments which likely
non-homologues included due to poor annotation, many involving {\it
  Eimeria tenella} (Et) sequences.

\begin{table}[!t]
\centering
\caption{Apicomplexa gene sets identified as outliers by \kdet{}. All annotations except 728 are putative.}
\label{names} 
  \begin{tabular}{lll}
    \hline
    No.$^a$ & GeneID$^b$ & Functional Annotation\\\hline
    488  &PF08\_0086 & RNA-binding protein\\
497  & PF13\_0228& 40S ribosomal subunit protein S6 \\
515  &PFA0390w&  DNA repair exonuclease\\
546 &PFF0285c &  DNA repair protein RAD50\\
547  & PFL1345c& Radical SAM protein\\
641  &PFE0750c&  hypothetical protein, conserved \\
 660 & PF10\_0043&  ribosomal protein L13 \\
662  & PF11\_0463& coat protein, gamma subunit \\
728 & MAL13P1.22& DNA ligase 1\\
747 & PFB0550w & Peptide chain release factor subunit 1 \\
773  & PFF0120w& geranylgeranyltransferase \\
780  & PFD0420c & flap exonuclease \\
\hline
\end{tabular}
\begin{flushleft}$^a$Based on geneset designations in \cite{kissinger}.\\
  $^b$Geneset represented by GeneID for \emph{Plasmodium falciparum}.
\end{flushleft} 
\end{table}


Since {\kdet{} revealed that there were} pervasive problems with the Et sequence
data, we removed this species from the data set and recreated the
phylogenetic analysis as in \cite{kissinger}. With the reduced set of
gene trees, \kdet{} identified a different set of outlier trees, and
in this case the Set8 gene was selected as the furthest outlying tree.

\subsubsection{Epichlo\"e:}
{The fungal datasets included alignments of ten fungal housekeeping
proteins, plus an alignment of suspected paralogs designated LolC and
CysD. The LolC/CysD tree was identified as one of two outliers,
the other being the DNA lyase protein ApnB. Topologically, the
LolC/CysD tree differed markedly from the others. However, the
topology of the ApnB tree was similar to the topologies for the other
housekeeping proteins, so its identification as an outlier suggested
that the ApnB 
tree had significantly different relative branch lengths from those of
the other housekeeping protein phylogenies in the analysis. }

\subsection{Running Time}
A significant advantage of \kdet{} over \pmc{} is a significant
improvement in computational speed, especially with larger data sets.
Actual \kdet{} running times are well fitted by a $O(N^2)$
curve, as suggested by the complexity of the algorithm discussed
previously, while the \pmc{} times are $O(N^3)$.

\section{Discussion}

 \subsection{Simulations}
 The results of our simulations were generally positive for \kdet{}. Although
 \pmc{} was often able to slightly outperform \kdet{} in classification
 accuracy, the difference was often relatively small. However, in terms of
 computational time, \kdet{} vastly outperforms \pmc{}, especially as
 the number of trees in the data set increases.

 In all cases studied, methods incorporating branch length information
 outperformed the topology only methods. The performance of the geodesic
 distance was better in the ``single'' simulations than the ``mixed''
 simulations, although the reason for this is unclear. All of the
 methods were able to correctly identify the outlier tree when the
 effective population size (and thus tree variance) was low, provided
 that a suitable tuning parameter was chosen. As the variance of the
 coalescent trees increased, the performance of \pmc{} tended to
 degrade at a slightly slower rate than \kdet{}.

 It should be noted that choosing a suitable tuning parameter can be
 quite difficult, as the optimal value depends on not only the details
 of the data set, but also one's subjective opinions on the relative
 merits of the sensitivity and specificity of the classifier. As such,
 we also studied the behavior of the algorithms when using their
 default tuning parameters. This information is relevant, since many
 users will not change the parameters from their default values. With
 these values we found that \kdet{} is slightly superior to \pmc{} in
 the single-distribution simulations. In the mixed-distribution
 simulations the default values for \pmc{} resulted in very poor
 performance, while \kdet{}'s rate of outlier identification was much
 higher.

 The third simulation set compared the distribution of scores for
 outlier trees to the scores of non-outlier trees.  Although the
 distributions are not completely distinct, it is clear that the
 outlier trees tend to have scores smaller than the majority of
 non-outlier trees. Since the outlier trees were generated as
 completely random coalescent trees, there will inevitably be trees
 generated which have structure similar to the non-outlier trees,
 simply by chance, and this accounts for some of the overlap between
 the distributions. With real data, such trees would correspond to
 genes which have some exotic history, but nonetheless appear to have
 a phylogeny substantially similar to the rest of the genes in the
 genome. In this case, it is ambiguous whether or not such a gene
 should be legitimately classified as an outlier.

 The main advantage of \kdet{} over \pmc{} lies in the vast
 improvement in running time on data sets with larger numbers of gene
 trees. For small data sets the difference is not material,
 however for data sets with several thousand trees, \pmc{} requires
 many hours to complete, while \kdet{} will finish within a few
 minutes on contemporary commodity hardware.

\subsection{Biological datasets}

\subsubsection{Apicomplexa:}

The phylum Apicomplexa contains many important protozoan pathogens
\citep{Levine1988}, including the mosquito-transmitted
\emph{Plasmodium} spp., the causative agents of malaria; \emph{T.
  gondii}, which is one of the most prevalent zoonotic pathogens
worldwide; and the water-born pathogen \emph{Cryptosporidium} spp.
Several members of the Apicomplexa also cause significant morbidity
and mortality in both wildlife and domestic animals. 
Due to their medical and veterinary importance, whole genome
sequencing projects have been completed for multiple prominent members
of the Apicomplexa.

The data set presented in \cite{kissinger} consists of $268$
orthologous genes from seven species of Apicomplexa and one outgroup
ciliate, \emph{Tetrahymena thermophelia}. To this set of genes we
appended sequences from the Set8 gene, which has been identified by
\cite{Kishore} as a probable case of horizontal gene transfer from a
higher eukaryote to an ancestor of the Apicomplexa.

While the Set8 gene was not identified initially by \kdet{} as an
outlier gene, its score was very close to the classification
threshold, and is the next gene to be classified as an outlier if the
tuning parameter is lowered slightly, from 1.5 to 1.3. Since many of
the outliers in the analysis seem to be caused by questionable
annotation in the Et sequences, we removed this species from the
data set and generated new gene trees. In the new analysis, the Set8
gene was identified as the furthest outlier tree.
These results demonstrate the potential applicability of the \kdet{}
method to the curation of genetic data sets by providing a simple tool
for highlighting sequences or alignments that may be of further
interest. The successful identification of the Set8 outlier indicates
that our method is able to highlight interesting cases which warrant
further attention from investigators.  {Moreover, the initial findings
with the Et sequences present in the dataset show that \kdet{} can be
useful for identifying problematic taxa due to incorrect annotation
and/or inclusion of non-orthologous genes.   }

\subsubsection{Epichlo\"e:}

The application of \kdet{} to the set of fungal protein alignments
successfully identified the paralogous CysD/LolC alignment as an
outlier. 
This is a scenario that could easily arise in
phylogenomic analysis, where OrthoMCL \citep{orthomcl} identified the
genes as orthologs, though the group was subsequently broken into
separate ortholog sets by application of COCO-CL \citep{cococl} to
the OrthoMCL output. 
{The identification of the LolC/CysD alignment as an outlier was
indicative of the utility of \kdet{} to identify outliers arising
from paralogy. }


\section{Conclusion}

The ongoing development of ever-cheaper sequencing methods is
producing a plethora of data suitable for phylogenomic analysis. One
of the great promises of modern genomics is that phylogenetics applied
at the genomic scale (phylogenomics) should be especially powerful for
elucidating gene and genome evolution, relationships among species and
populations, and processes of speciation and molecular
evolution. However, 
genomic data that can now be generated relatively cheaply and quickly,
but for which computationally efficient analytical tools are
lacking. There is a major need to explore new approaches to undertake
comparative genomic and phylogenomic studies much more rapidly and
robustly than existing tools allow.

In simulations and applications to biological data, we address
particular challenges posed by bioinformatic artifacts, as well as
interesting biological phenomena such as gene duplications and
horizontal gene transfer.  As we observed in the Apicomplexa and
fungal data sets, our approach also serves as a means of identifying
``interesting'' gene trees which may arise from horizontal gene
transfer, paralogy, or experimental artifacts such as misannotations
or misalignments.



A further advantage of our method is that it may be applied in a
straightforward way to phylogenetic reconstruction methods which
produce a a sample of many trees as output, rather than a single
``best fit'' tree. Indeed, methods that produce only a point estimate
does not represent the full set of possible phylogenies compatible
with the gene sequences.  We can circumvent this issue by building a
kernel for each gene based on a collection or sample of reconstructed
topologies (via the estimated posterior distribution of each gene, for
example), rather than using only a point estimate of each gene tree.

In future work we intend to extend our method to clustering trees
based on similarity, in addition to identifying outliers. The
identification and exclusion of outlier points is an important
preliminary step in many clustering methods.  The removal of outlier
points facilitates better inference at the clustering stage
\citep{Camastra2005,Hur2001,Hur2000}.

A long-term goal for this project is to develop a phylogenomic
pipeline that is convenient and accessible, as well as robust. To
accomplish this aim, important problems that need attention are (1)
refinement of gene calls based on comparison among orthologs from
multiple genomes and (2) comparing thousands of gene phylogenies
across whole genomes.  Therefore, our approach is focused on the
efficiency of the algorithm in terms of computational complexity and
memory requirements, with less emphasis on achieving the highest
classification accuracy possible.  Such a tradeoff makes our approach
more attractive candidate for inclusion in a pipeline for genome-wide
phylogenetics as an annotation supplement or as a discovery aid for
instances where evolutionary processes deviate significantly from
normal.


\section*{Acknowledgements}
We thank Drs.~Chih-Horng Kuo (Academia Sinica, Taiwan) and Jessica
Kissinger (Univ. of Georgia) for providing the Apicomplexa data set.
We would also like to offer special thanks to Dr.~de~Vienne, and our other
reviewers, for numerous valuable comments which have greatly improved
this manuscript.

\paragraph{Funding\textcolon} This work was supported by a grant from
the National Institute of Health [R01GM086888 to G.W., C.S., P.H., and
R.Y.]; and by a grant from the US Department of
Agriculture [USDA/CSREES award 2009-65109-05918 to D.H.].

\bibliographystyle{natbib}
\bibliography{gene}

\end{document}